# Multi-Agent Based Simulation for Investigating Centralized Charging Strategies and their Impact on Electric Vehicle Home Charging Ecosystem

Christensen, Kristoffer; Jørgensen, Bo Nørregaard; Ma, Zheng Grace

# Multi-Agent Based Simulation for Investigating Centralized Charging Strategies and their Impact on Electric Vehicle Home Charging Ecosystem

Kristoffer Christensen[1][0000-0003-2417-338X], Bo Nørregaard Jørgensen[1][0000-0001-5678-6602] and Zheng Grace Ma[1][0000-0002-9134-1032]

[1] SDU Center for Energy Informatics, Maersk Mc-Kinney Moeller Institute, The Faculty of Engineering, University of Southern Denmark, Odense, Denmark
`zma@mmmi.sdu.dk`

**Abstract.** This paper addresses the critical integration of electric vehicles (EVs) into the electricity grid, essential for achieving carbon neutrality by 2050. The rapid increase in EV adoption poses significant challenges to the existing grid infrastructure, particularly in managing the increasing electricity demand and mitigating the risk of grid overloads. Centralized EV charging strategies are investigated due to their potential to optimize grid stability and efficiency, compared to decentralized approaches that may exacerbate grid stress. Utilizing a multi-agent based simulation model, the study provides a realistic representation of the electric vehicle home charging ecosystem in a case study of Strib, Denmark. The findings show that the Earliest-deadline-first and Round Robin performs best with 100% EV adoption in terms of EV user satisfaction. The simulation considers a realistic adoption curve, EV charging strategies, EV models, and driving patterns to capture the full ecosystem dynamics over a long-term period with high resolution (hourly). Additionally, the study offers detailed load profiles for future distribution grids, demonstrating how centralized charging strategies can efficiently manage grid loads and prevent overloads.

## 1 Introduction

The evolution of the traditional electricity grid into a Smart Grid involves the integration of digital energy solutions to tackle the challenges brought about by the move towards carbon neutrality. This shift, in line with the European Union's objective of achieving a carbon-neutral society by 2050 [1], calls for the electrification of the transport sector. The swift transition from conventional vehicles to Electric Vehicles (EVs) is essential but requires a robust charging infrastructure to accommodate the growing electricity needs [2].

EV charging infrastructure is typically concentrated in areas with robust electrical grids, ensuring reliable access to power at large charging stations. However,

Distribution System Operators (DSOs) do not possess the regulatory authority to dictate whether individual residential consumers can install home charging stations. This regulatory gap poses a unique challenge, as DSOs are simultaneously tasked with guaranteeing high security of supply within a grid infrastructure originally designed for conventional energy consumption, devoid of the additional demands from EVs, photovoltaic systems, and heat pumps. Therefore, this paper focuses on home charging of EVs, highlighting its critical importance and exploring the implications for residential grid stability and energy management.

Digital solutions that exploit the flexibility of EV charging, such as intelligent EV charging systems and smart chargers, can significantly cut down the costs related to the integration of EVs into the electrical grids [3, 4]. Nevertheless, the increasing number of EVs threatens to cause significant grid overloads under the existing conditions. DSOs are faced with the decision of either expensive grid upgrades or the adoption of dynamic tariff schemes that leverage energy flexibility to manage and potentially postpone these upgrades.

Research and interviews with Danish DSOs reveal the challenge of the unpredictable outcomes of adopting EVs, and various charging strategies. These strategies aim to utilize the flexibility of EVs, but the overall effects on all ecosystem stakeholders—including DSOs and EV users—are not fully comprehended, making the strategic decisions to prevent economic losses and unnecessary expenditures complex.

Moreover, the introduction of dynamic electricity pricing structures, such as Time-of-Use tariffs and hourly electricity prices, is shaping consumers' awareness and consumption habits. These pricing models prompt consumers to be more mindful of their electricity usage times and volumes, thereby adding a layer of complexity to EV charging, which goes beyond the simple act of plug-in and charging. Before 2021 in Denmark, many consumers adopted a basic plug-in and charge approach, primarily due to a lack of awareness of dynamic pricing [5]. This strategy, known as traditional charging, is evolving post-2021 with the introduction of hourly price settlements, potentially changing consumer behavior.

Furthermore, smart charging, as defined by [6] and [7], entails the intelligent control of EV charging through connections between EVs, charging operators, and utility companies enabled by data links. Several EV charging strategies are proposed in the literature, e.g., [8-11]. Popular decentralized strategies are investigated by [12], finding that the investigated decentralized strategies induced earlier grid overloads compared to traditional charging. Decentralized charging strategies typically focus on EV users' preferences, such as minimizing charging costs by responding to price signals like day-ahead prices, which reflect overall grid conditions and encourage consumption during periods of lower demand.

However, these price signals often fail to consider local grid conditions within the distribution network, leading to synchronized charging behaviors that can overload local grids. In contrast, centralized charging strategies aim to mitigate this issue by controlling EV charging based on real-time local grid conditions. Despite the potential benefits, the future loading profile resulting from the implementation of technologically mature and user-friendly centralized charging strategies remains uncertain. These strategies are not limited by regulatory constraints and cause minimal inconvenience to EV users, making them crucial for DSOs to manage the increasing electricity demand effectively. Furthermore, the impact of these strategies on various stakeholders within the EV home charging ecosystem warrants comprehensive investigation. This paper addresses the urgent need to explore centralized EV charging strategies to enhance grid stability and optimize resource allocation within local grids.

This paper investigates the impact of technologically mature and user-friendly centralized EV charging strategies, as identified by [13]. The strategies under consideration include Round Robin, First-Come-First-Serve (FCFS), Equal Charge, and Earliest Deadline First (EDF), all considered feasible due to their current use at charging stations [14]. Using a multi-agent-based simulation model, this study explores the dynamics of these strategies in a case study involving 126 households in Strib, Denmark. This simulation approach allows for the representation of the ecosystem through interacting agents that mirror real-world interactions among ecosystem actors. By using real-world data on EV adoption rates, vehicle models, driving patterns, and base consumption, the simulation accurately reflects the EV home charging ecosystem [15]. The centralized charging strategies are compared against traditional charging methods to assess their impact on the ecosystem. The primary stakeholders examined in this paper are the Distribution System Operator (DSO) and EV users because the DSO is responsible for maintaining grid stability, while EV users are directly affected by the charging strategies.

The remainder of this paper is structured as follows: The Methodology section presents the multi-agent simulation framework, detailing the parameters and configurations employed for modeling centralized EV charging strategies. This is followed by the Case Study and Scenarios section, which applies these strategies within a specific context. Next, the Results and Discussion section evaluates these strategies in comparison to traditional charging methods. The paper concludes with the Conclusion section, summarizing key findings and implications.

## 2   Methodology

To study the impact of introducing centralized EV charging strategies on the EV home charging ecosystem, this paper developed a multi-agent-based simulation grounded in the identified business ecosystem. A business ecosystem comprises market actors, objects, and their respective roles. The complexity of this ecosystem arises from interactions between roles, including the flow of goods, monetary value, information, data, and intangible value [16]. The ecosystem framework used in this study is based on the EV home charging ecosystem identified by [17].

The translation to a multi-agent-based simulation follows the methodology presented in [18], where ecosystem actors and objects are converted into agents with roles integrated through Java interfaces or abstract classes to ensure agents include the responsibilities and functionality of the assigned roles. This approach is essential because analytical models are often inadequate for addressing complex problems characterized by dynamic systems with non-linear behavior, memory effects, intricate variable interdependencies, temporal factors, and cause-and-effect relationships [15]. Such properties render analytical models unsuitable for realistic energy ecosystems due to its stochastic elements and events with different triggers such as time-based.

Agent-based modeling, chosen for this study, simulates the interactions of individual agents to assess their effects on the system as a whole. This method allows for a detailed and dynamic analysis of both the overall ecosystem (high abstraction) and the specific behaviors of individual consumers and electric vehicles (low abstraction). This dual-level analysis is particularly valuable for understanding the multifaceted nature of EV home charging ecosystems and the impact of centralized charging strategies. Among several agent-based modeling tools, the multi-method simulation tool – AnyLogic [19], has been selected based on its application domains within power grids, business strategy, and innovation analysis.

The agents found relevant to address for simulation experiments conducted in this paper are the DSO, Charging box, EV, Domestic consumer, Charging service provider, and centralized charging strategies which are modeled as agents for enhanced modularity in the model. The logic of the different strategies is as follows:

- Round Robin: Charges EVs in intervals, e.g., an EV starts charging and exceeds the maximum allowed number of simultaneous charging EVs. The newly arrived EV will not charge before the next time interval starts. When the new interval starts, the newly arrived EV will charge, and one of the already charging EVs will stop charging. The next interval will then stop the charging at another EV instead, and so it continues until the number of simultaneous charging EVs is below the allowed limit [14].
- FCFS: Charges the EVs in the order they arrive. For instance, an EV arrives at home and connects to the charging box which requests charging permission from the centralized management system managed by the charging service provider. If the charging of the EV exceeds the capacity, the EV is put first in the queue for charging when capacity is sufficient [20].
- Equal Charge: Shares the available capacity equally between all connected EVs up to their maximum charging rate [14]. When the maximum charging rate of all connected EVs exceeds the limit, the charging rate is reduced so that all EVs charge at the same rate but with a total rate that is below the limit. The strategy takes into account if EVs charge at different rates. For instance, this strategy does not affect EVs charging with 3.7 kW if the capacity limit is satisfied by reducing EVs with an 11 kW charging rate to anything higher than 3.7 kW.
- EDF: Charges always the EV with the earliest departure time first [10], with the assumption that EV users provide their planned departure time.

## 3      Case study and simulation experiments

A case study of EV home charging in a distribution network in Strib, Denmark, consisting of 126 households is used. The consumption data was supplied by the Danish Distribution System Operator (DSO) TREFOR, which is responsible for this region. The data utilized is solely for 2019, deliberately omitting the years 2020-2021 and 2022 which were affected by the COVID-19 pandemic and the energy crisis due to the war between Russia and Ukraine, respectively. The case study's low voltage electricity network (400 V) is connected to a 0.4/10 kV transformer with a capacity of 400 kVA. This system is simulated for residential households, with an expected power factor close to one, and a transformer capacity of 400 kW. For simplicity, the transformer's capacity is taken as the total capacity of the distribution network, disregarding the cables and nodes within the network. The EV adoption is a Poisson process based on the adoption curve estimated in [15] and shown in Fig. 1, achieving 100% adoption within 2039.

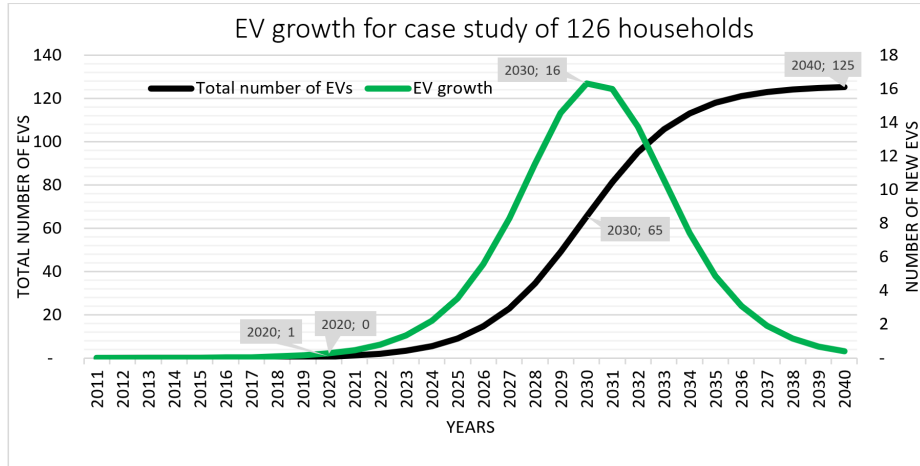

**Fig. 1.** Adoption curve (black) and electric vehicles adopted per year (green) [15].

The driving behavior is based on the consumption pattern found in the baseload. When it comes to choosing specific EV models, the decision is made randomly, with probabilities based on the popularity of the top five EV models in Denmark during 2019. More details are explained in [15].

To investigate the ecosystem impact of the proposed centralized strategies, 10 experiments are designed and listed in Table 1. Each experiment is based on one of the four centralized strategies (Round Robin, Equal Charge, FCFS, and EDF) with an hourly electricity price scheme and fixed or dynamic distribution tariff. The Round Robin strategy is the only dependent on a time interval between cycles, here 15 minutes are chosen. To evaluate the effect of the centralized strategies, the traditional charging strategy - i.e. charging immediately at arrival, is used as a baseline for comparison with the centralized strategies.

**Table 1.** Simulation experiments.

| Number | Experiment |
|---|---|
| 1 | Traditional charging |
| 2 | Traditional charging for 2036 to 2039 |
| 3 | Round Robin |
| 4 | Round Robin for 2036 to 2039 |
| 5 | Equal Charge |
| 6 | Equal Charge for 2036 to 2039 |
| 7 | First-Come-First-Serve |
| 8 | First-Come-First-Serve for 2036 to 2039 |
| 9 | Earliest Deadline First |
| 10 | Earliest Deadline First for 2036 to 2039 |

The simulation was conducted from 2020 to 2032, as it was found in [12] that the decentralized charging strategies induced overload within this period. However, for each of the four centralized strategies, an additional experiment for 2036 until 2039 is designed to investigate their impact on a system with 100% EV adoption. In 2032, 89-98 EVs are adopted, corresponding to 71-78% adoption share. To evaluate the charging strategies the following key performance indicators are chosen: Overloads the year after the first overload; average charging cost of EV users; average total electricity bill (includes all electricity consumption) of all EV users; average total CO2 emissions from consuming electricity; EV users' dissatisfaction (i.e. desired charging level is not reached); Load factor of the grid; and DSO revenue from distribution tariff alone. The mentioned performance indicators are compared between the different strategies, including the traditional charging, i.e. starting the charge at arrival.

## 4    Results

- **Traditional charging**

The results of traditional charging from 2020 to 2032 are presented in [15]. [15] does not simulate the traditional charging strategy for 2036 to 2039. Therefore, to compare the centralized strategies performance with 100% EV adoption, traditional charging is simulated from 2036 to 2039.

Table 2 shows the key results for the traditional charging, from 2036 to the last simulation year (2039). Overload occurs on the first day of the simulation period, and overloads occur almost daily due to the 100% EV adoption.

**Table 2.** Key results for the traditional charging, from 2036 to 2039. The results are calculated for the last simulation year (2039).

| Experiment | Overloads the year after the first overload | Avg. charging cost for an EV user [DKK] | Avg. total electricity bill [DKK] | Avg. total CO2 emissions [kg] | EV users' dissatisfaction | Load factor | DSO revenue [DKK] |
|---|---|---|---|---|---|---|---|
| Traditional charging | 543 | 1.3495 | 11,025.08 | 589.4865 | 59 | 0.2025 | 168,397.66 |

Fig. 2 shows the total grid load for 2036 to 2039 using the traditional charging strategy. Fig. 3 shows the EV users' dissatisfaction i.e. whenever the EV is not charged to the desired state-of-charge. Each EV user is listed on the primary axis, showing the number of dissatisfactions. Only users with a Nissan Leaf experience dissatisfaction due to its low charging rate of 3.7 kW and situations when they arrive late with a low state-of-charge and early departure. Except for the Equal Charge, for all centralized strategies, the dissatisfaction is the same as for this experiment and will not be discussed further.

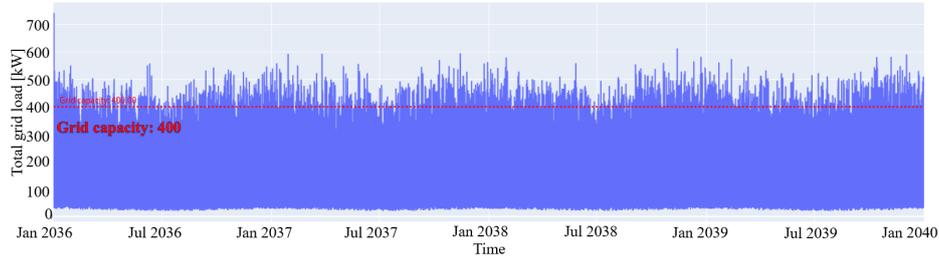

**Fig. 2.** Total grid load for traditional charging strategy from 2036 to 2039.

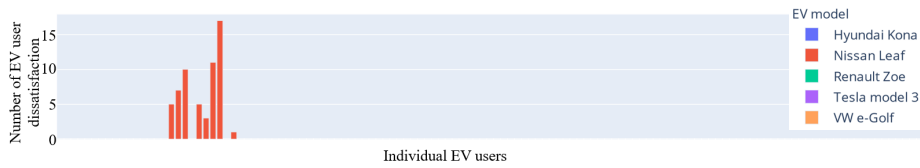

**Fig. 3.** Electric vehicle users' dissatisfaction with the traditional charging strategy in 2039.

### 4.1 Round Robin Charging Strategy

The logic of the Round Robin charging strategy is to keep the total load below transformer capacity by not having more EVs charging than the grid can manage. The Round Robin algorithm cycles the charging and non-charging queues for each time interval.

- **Round Robin for 2020 to 2032**

Table 3 shows the key results from Round Robin compared to the traditional charging.

**Table 3.** Key results for the Round Robin charging strategy compared with traditional charging.

| Experiment | Avg. charging cost for an EV user [DKK/kWh] | Avg. total electricity bill [DKK] | Avg. total $CO_2$ emissions [kg] | Load factor | DSO revenue [DKK] |
|---|---|---|---|---|---|
| Round Robin | 1.3714 | 11,229.52 | 53.4904 | 0.252 | 140,975.35 |
| Traditional charging | 1.3714 | 11,229.9 | 53.4965 | 0.2048 | 140,998.34 |
| Percentage difference | 0.00% | 0.00% | -0.01% | 23.05% | -0.02% |

Besides the load factor, no real difference is found in applying the Round Robin charging strategy to the system. It can be concluded that the Round Robin can be used until the year 2032 to avoid overload and better utilize grid capacity (due to the increased load factor) without affecting EV users.

Fig. 4 shows the total grid load from 2031 to 2032. The figure shows that the load stays below the grid capacity. The dissatisfactions from the Round Robin strategy are equal to the traditional charging.

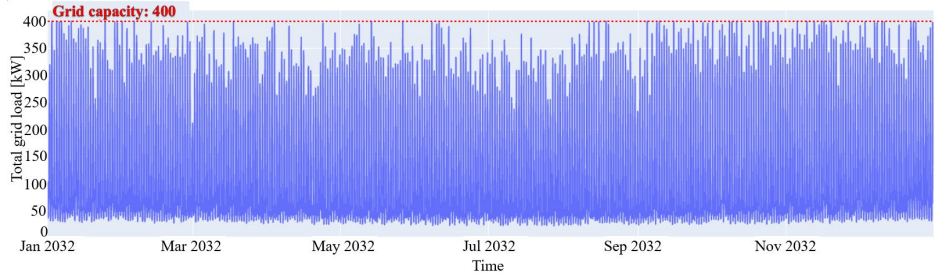

**Fig. 4.** Total grid load example.

- **Round Robin for 2036 to 2039**

This experiment aims to test how the Round Robin charging strategy performs under 121 to 126 EVs in the grid (96-100% EV adoption). The relevant results of this experiment are to see if this charging strategy can avoid overload with 100% EV adoption and if any EV user (except the Nissan Leaf) experiences any dissatisfaction. The key result for 2039 of this experiment is shown in Table 4 and compared with traditional charging, from 2036 to 2039.

**Table 4.** Key results for 2039 of the Round Robin 2036 to 2039 experiment, compared with the result of traditional charging 2036 to 2039 experiment. (Note the results are for 2039)

| Experiment | Avg. charging cost for an EV user [DKK/kWh] | Avg. total electricity bill [DKK] | Avg. total $CO_2$ emissions [kg] | EV users' dissatisfaction | Load factor | DSO revenue [DKK] |
|---|---|---|---|---|---|---|
| Round Robin | 1.3482 | 11,018.75 | 588.306 | 62 | 0.2977 | 167,742.34 |
| Traditional charging | 1.3495 | 11,025.08 | 589.4865 | 59 | 0.2025 | 168,397.66 |
| Percentage difference | -0.10% | -0.06% | -0.20% | 5.08% | 47.01% | -0.39% |

Table 5 shows the total grid load between 2036 and 2039. The figure shows that the Round Robin keeps the load below grid capacity. No EV user (except the Nissan Leaf) experiences dissatisfaction.

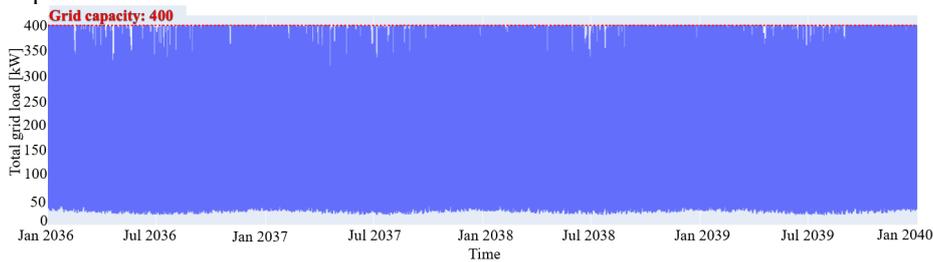

**Table 5.** Total grid load for Round Robin from 2036 to 2039.

### 4.2 Equal Charge Strategy

- **Equal Charge for 2020 to 2032**

Table 6 shows the results for the Equal Charge compared with the traditional charging.

The results show that, like the Round Robin, the Equal charge has no real impact on the system except for the increased load factor. However, the DSO's revenue is slightly more affected than the Round Robin strategy.

**Table 6.** Key results for the Equal charge charging strategy compared with traditional charging.

| Experiment | Avg. charging cost for an EV user [DKK/kWh] | Avg. total electricity bill [DKK] | Avg. total $CO_2$ emissions [kg] | Load factor | DSO revenue [DKK] |
|---|---|---|---|---|---|
| Equal Charge | 1.3715 | 11,230.29 | 53.5006 | 0.252 | 140,472.88 |
| Traditional charging | 1.3714 | 11,229.9 | 53.4965 | 0.2048 | 140,998.34 |
| Percentage difference | 0.01% | 0.00% | 0.01% | 23.05% | -0.37% |

- **Equal Charge for 2036 to 2039**

The key results for the Equal Charge charging strategy in 2039 with a comparison of the results of the traditional charging from 2036 to 2039 experiment are shown in Table 7.

**Table 7.** Key results for the Equal Charge, from 2036 to 2039 experiment compared with traditional charging. (Note the results are for 2039)

| Experiment | Avg. charging cost for an EV user [DKK/kWh] | Avg. total electricity bill [DKK] | Avg. total $CO_2$ emissions [kg] | EV users' dissatisfaction | Load factor | DSO revenue [DKK] |
|---|---|---|---|---|---|---|
| Equal charge | 1.3325 | 10,938.36 | 580.6725 | 97 | 0.2977 | 147,006.17 |
| Traditional charging | 1.3495 | 11,025.08 | 589.4865 | 59 | 0.2025 | 168,397.66 |
| Percentage difference | -1.26% | -0.79% | -1.50% | 64.41% | 47.01% | -12.70% |

Like the Round Robin, the equal charge avoids overload with a 100% EV adoption. The Equal Charge performs the best on the EV users' charging costs and electricity bills. However, as shown in Fig. 5, it results in more dissatisfaction; ninety-seven dissatisfactions during 2039, 64% more than the traditional charging. For the EV users (not owning a Nissan Leaf), the dissatisfaction happens mainly once, meaning that it happens on rare occasions (none for Renault Zoe users). Therefore, users with fast-charging EV models will benefit from the Equal Charge strategy. However, the Equal Charge significantly impacts the DSO's revenue with a reduction of almost 13%, whereas the Round Robin is only a 0.39% reduction.

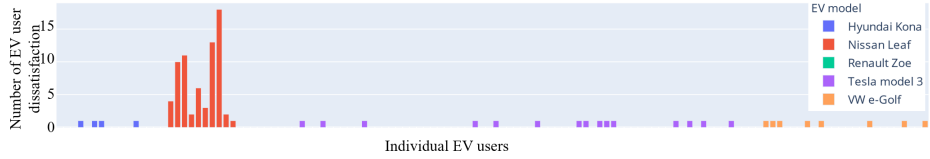

**Fig. 5.** Electric vehicle users' dissatisfaction with Equal Charge charging strategy in 2039.

### 4.3    First-Come-First-Serve Strategy

- **First-Come-First-Serve for 2020 to 2032**

Table 8 shows the experiment results for using FCFS with a comparison with traditional charging.

The experiment results of using FCFS are very similar to the experiment results of using Round Robin and Equal Charge strategies. However, the DSO's revenue is less affected than the Equal Charge strategy.

**Table 8.** Key results for the First-Come-First-Serve with a comparison with traditional charging.

| Experiment | Avg. charging cost for an EV user [DKK/kWh] | Avg. total electricity bill [DKK] | Avg. total $CO_2$ emissions [kg] | Load factor | DSO revenue [DKK] |
|---|---|---|---|---|---|
| First-Come-First-Serve | 1.3714 | 11,229.51 | 53.4905 | 0.252 | 140,974.88 |
| Traditional charging | 1.3714 | 11,229.9 | 53.4965 | 0.2048 | 140,998.34 |
| Percentage difference | 0.00% | 0.00% | -0.01% | 23.05% | -0.02% |

- **First-Come-First-Serve for 2036 to 2039**

The key experiment results for using FCFS in 2039 and the comparison results with the traditional charging experiment are shown in Table 9. FCFS avoids overload with a 100% EV adoption. The FCFS performs almost as well as Round Robin except for a slight difference in user dissatisfaction (FCFS has one more) and the DSO revenue (FCFS has around 38 DKK less).

**Table 9.** Key results for the First-Come-First-Serve from 2036 to 2039 experiment, compared with traditional charging. (Note the results are for 2039).

| Experiment | Avg. charging cost for an EV user [DKK/kWh] | Avg. total electricity bill [DKK] | Avg. total $CO_2$ emissions [kg] | EV users' dissatisfaction | Load factor | DSO revenue [DKK] |
|---|---|---|---|---|---|---|
| First-Come-First-Serve | 1.3482 | 11,018.27 | 588.2826 | 63 | 0.2977 | 167,704.32 |
| Traditional charging | 1.3495 | 11,025.08 | 589.4865 | 59 | 0.2025 | 168,397.66 |
| Percentage difference | -0.10% | -0.06% | -0.20% | 6.78% | 47.01% | -0.41% |

### 4.4 Earliest Deadline First Strategy

- **Earliest Deadline First for 2020 to 2032**

Table 10 shows the experimental results of using EDF compared to the traditional charging. The results show that the EDF charging strategy's performance equals the Round Robin and FCFS for 2032.

**Table 10.** Key results for the Earliest Deadline First compared to the traditional charging.

| Experiment | Avg. charging cost for an EV user [DKK/kWh] | Avg. total electricity bill [DKK] | Avg. total $CO_2$ emissions [kg] | Load factor | DSO revenue [DKK] |
|---|---|---|---|---|---|
| Earliest Deadline First | 1.3714 | 11,229.59 | 53.4905 | 0.252 | 140,975.24 |
| Traditional charging | 1.3714 | 11,229.9 | 53.4965 | 0.2048 | 140,998.34 |
| Percentage difference | 0.00% | 0.00% | -0.01% | 23.05% | -0.02% |

- **Earliest Deadline First for 2036 to 2039**

The key experiment results for using the EDF charging strategy with a 100% EV adoption in 2039 are shown in Table 11 with a comparison of the traditional charging from 2036 to 2039 experiment.

**Table 11.** Key results for the Earliest Deadline First from 2036 to 2039 experiment, compared with traditional charging. (Note the results are for 2039).

| Experiment | Avg. charging cost for an EV user [DKK/kWh] | Avg. total electricity bill [DKK] | Avg. total $CO_2$ emissions [kg] | EV users' dissatisfaction | Load factor | DSO revenue [DKK] |
|---|---|---|---|---|---|---|
| Earliest Deadline First | 1.3483 | 11,019.18 | 588.3638 | 59 | 0.2977 | 167,770.54 |
| Traditional charging | 1.3495 | 11,025.08 | 589.4865 | 59 | 0.2025 | 168,397.66 |
| Percentage difference | -0.09% | -0.05% | -0.19% | 0.00% | 47.01% | -0.37% |

The EDF is the only centralized strategy having the same number of dissatisfactions as without any centralized strategies. The reason is that the users with the earliest departure are prioritized and require that the algorithm knows when users are planning to leave. Besides this, the EDF strategy performs slightly more poorly than the Round Robin and FCFS in the other key results, except for the DSO's revenue, where the EDF performs slightly better.

## 5  Discussion

The primary objective of centralized EV charging strategies is to prevent grid overloads. The results demonstrate that overloads can be effectively avoided with approximately 98 EVs in the grid until the end of 2032. Up to an 80% EV adoption rate, the implementation of any of the four centralized strategies by the DSO or a third party does not significantly affect EV users, assuming they follow traditional charging behaviors. This highlights the robustness of centralized strategies in managing increased loads without compromising user satisfaction.

### 5.1  Example of Round Robin Strategy

Centralized strategies effectively manage EV charging without compromising user satisfaction, as illustrated by the Round Robin strategy example. Fig. 6 shows the total grid load for a day in 2032 experiencing overload under traditional charging and the same day using the Round Robin strategy. The Round Robin strategy prevents overloads (e.g., on January 29 at 5 and 6 PM) without causing a significant load increase in the following hours. This is because the simulation output is hourly-based, showing the maximum load within each hour. Consequently, EVs are charged within the overload hours, as the load does not increase in subsequent hours. Table 12 further supports this by detailing the number of charging EVs and overload durations for traditional and Round Robin charging experiments:

- At 5 PM, there are 18 minutes of overload within the hour, with a maximum overload size of 74 kW. This implies that 42 minutes within the hour are free from overload, indicating available capacity. However, due to the overload size, it is not feasible to charge the same number of EVs within this hour using the Round Robin strategy, resulting in seven fewer charging EVs.
- At 6 PM, the entire hour experiences overload, leaving no room for load shifting. The Round Robin algorithm mitigates this by cycling new EVs to charge every 15 minutes while others are turned off, avoiding overload with only three fewer charging EVs.
- At 7 PM, no overload occurs, but the load is only 5 kW below grid capacity. The Round Robin strategy continues to manage charging to avoid overload, allowing two more EVs to charge during this hour.
- At 8 PM, the number of charging EVs remains consistent, suggesting that EVs charge for a longer duration within this hour under the Round Robin strategy. The number of charging EVs gradually increases by 1-2 for the next three hours, eventually matching the traditional charging pattern from 12 AM onward (hours after 12 AM are not shown as they are identical).

This example demonstrates that the Round Robin strategy only load-shifts a total of seven charging EVs within whole hours. Due to the available capacity within hours, other centralized strategies perform similarly, explaining the negligible impact on key results. Additionally, the load factor is significantly increased by 23%, indicating that grid capacity is utilized more efficiently with centralized strategies. This increase in load factor is consistent across all centralized strategies, making it challenging to determine the superior strategy among them.

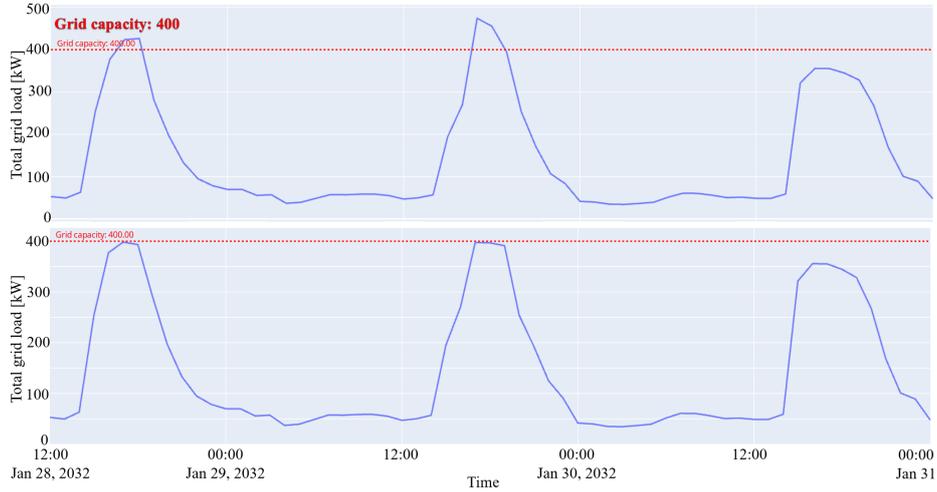

**Fig. 6.** Total grid load for the traditional charging (top) and for the Round Robin (bottom), an example of overload on January 29.

**Table 12.** Number of charging electric vehicles for traditional charging and Round Robin experiment.

| Hour | Traditional charging | Round Robin | Difference in number of charging EVs | Minutes with overload | Size of overload |
|---|---|---|---|---|---|
| 16 | 34 | 34 | 0 | - | - |
| 17* | 46 | 39 | -7 | 18 | 74.16 kW |
| 18* | 47 | 44 | -3 | 60 | 55.27 kW |
| 19 | 34 | 36 | +2 | - | - |
| 20 | 23 | 23 | 0 | - | - |
| 21 | 14 | 16 | +2 | - | - |
| 22 | 11 | 13 | +2 | - | - |
| 23 | 6 | 7 | +1 | - | - |
| 0 | 2 | 2 | 0 | - | - |

*Hours with overload in the traditional charging experiment

## 5.2 Assumptions and Practical Considerations

The centralized EV charging algorithms assume real-time knowledge of changes in baseload. In our simulation, baseload information is updated simultaneously for both centralized strategy agents and consumer agents, implying no delay. However, in reality, even a microsecond delay could result in an overload. Therefore, incorporating a buffer within the algorithm, combined with AI-based load forecasting, is necessary. The buffer can be determined using historical load data, allowing the algorithms to operate at the transformer capacity minus the buffer.

It is important to note that the centralized EV charging strategies investigated in this research are primarily designed for charging stations. Consequently, their practical application in home charging systems may be limited. For example, the EDF strategy might be unsuitable for distribution grids, as users could set early departure times to gain charging priority (unless the centralized strategy is unknown to users). Additionally, the fairness of centralized EV charging strategies for individual users (e.g., those arriving late in FCFS or departing late in EDF, and those with low charging rates) warrants further investigation. Future research should explore how these strategies can be adapted for home charging scenarios and assess their fairness and efficiency in diverse real-world contexts.

## 6      Conclusion

The rapid increase in electric vehicle (EV) adoption poses significant challenges for Distribution System Operators (DSOs) tasked with ensuring reliable electricity supply through grids not originally designed for such loads. To address this issue without incurring substantial grid reinforcement costs, it is essential to explore cost-effective, technologically mature and user-friendly centralized EV charging strategies that maximize the current grid capacity.

This paper investigates the effectiveness of centralized EV charging strategies—Round Robin, First-Come-First-Serve (FCFS), Equal Charge, and Earliest Deadline First (EDF)—in preventing grid overloads. By employing a multi-agent-based simulation, the study models the EV home charging ecosystem with realistic adoption curves, diverse charging strategies, various EV models, and driving patterns over a long-term period with high temporal resolution (hourly).

The findings indicate that all examined centralized charging strategies successfully prevent grid overloads, a critical consideration as traditional and decentralized strategies often lead to earlier grid overloads. At an 80% EV adoption rate, centralized strategies exhibit a minimal impact on user satisfaction. However, when EV adoption reaches 100%, the EDF strategy proves most effective in minimizing user dissatisfaction. Despite its efficacy, practical implementation of the EDF strategy could lead to strategic manipulation by users, suggesting that the Round Robin strategy, which ranks second in reducing dissatisfaction, may be a more pragmatic choice.

The study provides significant contributions by offering detailed future load profiles of distribution grids under various centralized charging strategies, demonstrating their practical viability in preventing grid overloads and maintaining user satisfaction. These findings have profound practical implications, equipping DSOs with actionable strategies to manage the increasing electricity demand due to EV adoption. By implementing these centralized strategies, DSOs can effectively defer costly grid reinforcements and ensure a stable and reliable electricity supply.

The main limitation of the study is the assumption that users strictly adhere to pre-defined charging behaviors, without accounting for potential behavioral adaptations over time. Additionally, the focus on a specific geographic area and set of conditions may limit the generalizability of the results. Future research should address these limitations by incorporating adaptive user behaviors, exploring diverse geographic contexts, and considering the impact of regulatory changes. Moreover, examining the combined effects of decentralized and centralized strategies will provide a more comprehensive understanding of their potential to optimize grid performance and user satisfaction.

While centralized EV charging strategies show promise in preventing grid overloads and improving grid capacity utilization, further research is needed to address practical implementation challenges, user behavior adaptations, and the fairness of these strategies across different user scenarios. This will ensure the development of robust, adaptable, and equitable solutions for managing the increasing electricity demand associated with EV adoption.

**Acknowledgments.** This research is part of the Digital Energy Hub funded by the Danish Industry Foundation and the IEA ES Task 43 "Storage for renewables and flexibility through standardized use of building mass", funded by EUDP (case number: 134232-510227).

**Disclosure of Interests.** The authors have no competing interests to declare that are relevant to the content of this article.